\documentstyle[preprint,aps]{revtex}
\begin{document}

\title{
%\begin{flushright}
%\vspace{-1cm}
%{\normalsize MC/TH 96/19}
%\vspace{1cm}
%\end{flushright}
Chiral dynamics of the low energy \\
 Kaon-Baryon interactions.}
\author{ Boris Krippa$^{**}$}
\address{Nuclear Theory Center, Indiana University,\\
2401 Milo Sampson Lane, Bloomington, IN 47408, USA.\\}
\maketitle
\begin{abstract}
The processes involving $K^{-}p, KN, \Sigma\pi, \Lambda\pi,\Sigma\eta,
 \Lambda\eta$ coupled channels are studied in the nonperturbative chiral approach.
 An effective potential is constructed using a chiral meson-baryon Lagrangian at lowest
 order. This potential is iterated to all orders with the Lippmann-Schwinger equation.
 A reasonable fit of the experimental
data is obtained. It is pointed out, however, that due to a strong sensitivity of the
 results to the value of the cut-off, such an approach should  be viewed as a rather
phenomenological way to fit the experimental data since there is no small expansion 
parameter allowing for truncation of the chiral expansion of the effective 
potential at some given order. A possible way to construct the consistent chiral
expansion is briefly discussed.\\ 
Pacs: (13.75.Gx; 12.38.Lg; 11.55.Hx; 14.20.Dh)
\vskip4cm 
\noindent $^{**}${\it On leave from the Institute for Nuclear Research 
of Russian Academy of Sciences, Moscow, 117312, Russia.} 

\end{abstract}

\newpage
Chiral perturbation theory (ChPT) has become at present a standard tool to study 
the low-energy hadronic interaction  \cite{ec}. ChPT is based on the idea 
of constructing the most general set of effective lagrangians consistent with the
symmetries of QCD and corresponding to the expansion of the scattering amplitudes
in increasing powers of mesonic momenta and/or quark masses. An important
 feature of ChPT is the chiral counting scheme which allows one to estimate the
 chiral dimension of the scattering amplitude at any given order, endowing the
 method with more predictability. Initially 
developed for the meson-meson interaction, ChPT 
was then generalized to include baryons in a fully relativistic manner  \cite{gss}.
Unfortunately, in this treatment the consistent chiral counting is difficult
 to implement, since the baryon 4-momenta can never be small in the scale, as is
typical for the low-energy meson-meson interactions. In ref.\cite{jm} a
 nonrelativistic formulation of ChPT was proposed, which allows one to reconstruct 
the power counting rules and thus organize the consistent chiral expansion
scheme. This Heavy Baryon Chiral Perturbation Theory (HBChPT), where baryons are assumed 
to be infinitely heavy in lowest order, has been 
successful \cite{mbk} in describing many properties of meson-baryon systems
at low energies. However, when applied to the NN  interaction
ChPT encounters a problem due to the existence of a bound state with an energy
close to threshold. This state cannot be reproduced by the standard means
of ChPT \cite{we}.
 Physically, it is rather natural, since in a bound state a particle interacts
an infinite number of times, whereas in a perturbative series only a finite
number of interaction is taken into account. Weinberg \cite{we} first suggested
the formulation of ChPT in nuclear physics, based on the idea that  power
counting arguments  can be applied to the NN (or n-nucleon) potential
instead of the scattering amplitude. This potential  can then be used to 
calculate, for example, NN phase shifts by iterating a Lippmann-Schwinger
or Schrodinger equation in all orders. However, as it was pointed out  
in Ref.\cite{bpc}, when the power counting scheme is used for the effective
 potential, which should then be iterated in all orders, some other problems
arise, making the consistent implementation of the above mentioned program rather
difficult. The results of the Ref. \cite {bpc} can be summarized as follows.
Firstly, it was shown  that there is no small parameter
 in the chiral
expansion of the effective potential, allowing for the truncation 
of the series at some given order. Secondly, it was demonstrated that different 
regularization schemes lead to different physical amplitudes. That means
a sensitivity to the short-range physics, that cannot be properly treated
in the framework of standard ChPT. These results, obtained for the
 case of NN scattering, indicate the necessity to consider the other hadronic
systems which are dominated by a bound state (or resonance) close to threshold.
In this paper we concentrate on the study of the $K^{-}p$ interaction.
The low energy properties  of the $K^{-}P$ system are, to large extent,
 determined by the $\Lambda$(1405) resonance, just below $K^{-}P$ threshold.
Since baryons in such an approach can be treated as ``heavy'' compared to
mesons one may hope
that the sensitivity to the cut-off parameter will not be so strong and a
 consistent chiral approach can be formulated. Below we show that 
 this is not the case. One notes, that  the $K^{-}p$ system 
was earlier considered in Refs.  \cite{kzw} and\cite{lee}. 
 In contrast to the $\pi N$ and
 $K^{+}N$ cases, the  $K^{-}P$ system is a strongly interacting system with
many possible channels, so it is reasonable to choose the scheme of calculations
in the spirit of that, suggested in  \cite {kzw}, where the effects of coupled
channels have been incorporated. We, however, have used an approach, which is
 somewhat different from that proposed in  \cite {kzw} and, in our opinion,
technically easier. That scheme was first developed and successfully
implemented in the case of meson-meson interactions in Ref.  \cite{oo}.
We start from the lowest order meson-baryon effective Lagrangian 
\begin{equation}
  {\cal L}^{(1)}_{int} = {i \over {4f^2}} Tr(\overline{B}[[\phi,
                                       \partial_0 \phi],B])
\label{twopt}
\end{equation}
where $B$ and $\phi$ are baryon and meson field matrices in their standard
 form  \cite {mbk}. After straightforward calculations one can show
 that in the lowest order the tree level amplitude has a form
\begin{equation}
t \simeq {1\over f_{\pi}^{2}}  (E_i - E_f)
\end{equation}
where $f_{\pi}$ is the pion decay constant, and $E_i(E_f)$ is the total energy
of the incoming (outgoing) mesons. According to the approach, outlined above,
this tree level amplitude should be substituted in the Lippmann-Schwinger
 equation for the full $T$-matrix
\begin{equation}
T=V+VGT
\end{equation}      
where V is the effective potential, which in our case is given by the tree
level amplitude $t$ and $G$ is the corresponding propagator, defined by 
\begin{equation}
  G_{l} = (2\pi)^{-4}
             \int{d^{4}q {i\over {(q^2 - m^2_l + i \epsilon)(p_0-q_0)}}},
\end{equation}  
where  $q_0$($m_l$) is the energy (mass) of the intermediate meson of the
type $l$. In Ref.\cite{oo}  it was demonstrated that one can avoid the 
problem of finding the complete solution of the Lippmann-Schwinger equation with 
off shell scattering amplitudes since, at least at leading order, only the on-shell
information about the tree level amplitudes is really needed. This procedure can easily
be extended for the case of meson-nucleon interaction in HBChPT.
 We consider for simplicity only the one-loop contribution with a cut-off $\Lambda$
imposed. The corresponding loop integral consists of meson and baryon 
propagators and two tree-level amplitudes. The typical expression for the numerator 
in the one loop term can be decomposed as follows
\begin{equation}
(p^0 +q^0)^2=4p^{0^{2}} + 4p^{0}(q^0 - p^0) + (p^0 -q^0)^2.
\end{equation}  
Putting this expression in the integral part of the Lippmann-Schwinger equation
 one can represent the loop corrections by the sum of three integrals. 
 First one gives the part of the loop corrections where the rescattering amplitude
 is taken on-shell. This integral will be calculated below. The other two integrals contain
 the contributions from the off-shell part of the rescattering amplitude. The expression
 $4p^{0}(q^0 - p^0)$, when substituted in the loop integral cancels the baryon propagator. One 
 can see that after the integration over energy is done the integral becomes proportional to 
$p^{0}\int d^3q (2\omega_l(q))^{-1} \simeq p^{0} \Lambda^{2}$, which has the
structure of the tree level amplitudes and can be combined with them.
A similar  property holds for the third term in Eq.(5)
and for the terms of higher order in the Lippmann-Schwinger equation.
Thus, the corresponding integral equation can be reduced to a system of
algebraic equations and the amplitude $T$ is given by   

\begin{equation}
T=(1 - VG)^{-1}V
\end{equation} 
One notes that the presence of cut-off parameters is inevitable in an 
approach of this type. In the ordinary ChPT the divergencies can be removed
order by order, but since we have to sum the series to all orders, some
regulator should always present. One can still formulate a consistent chiral
expansion scheme if the value of the cut off parameter agrees with
the chiral symmetry expectations and the dependence of the physical observable
on the value of the cut-off parameter is moderate.

 There are eight channels ($\pi\Sigma, \pi\Lambda, K^{-}p, KN,\eta\Lambda,
\eta\Sigma$)
 which can be involved in the dynamics
of the $K^{-}P$ interaction at low energies. 
After imposing the cut off and taking the tree level amplitude out
of the integral in the Lippmann-Schwinger equation the only remaining task
is to calculate the residual integral. It can be done analytically, using
 standard field theoretical methods. The result of the calculations
is given in the appendix.

 The results of our calculations for  elastic scattering
 $K^{-}p \rightarrow K^{-}p$, and for the reactions  $K^{-}p \rightarrow  \pi\Sigma$
 are shown in Fig.1-3.
In Figs.1-2 the lower lines correspond to calculations using the value of
the cut off parameter  $\Lambda$=600 MeV, whereas the upper curves 
are the results of calculations with $\Lambda$=580 MeV. The theoretical
calculations are in reasonable agreement with the experimental data. We note that, 
in principle, one can take into
 account the baryon mass splitting in the SU(3) multiplets when calculating
the propagator $G_l$. The results of such calculations are shown in Fig.3
for the reaction  $K^{-}p \rightarrow  \pi^{-}\Sigma^{-}$. At 
threshold the total cross section gets further enhanced. It could be viewed as a
hint of the existence of the resonance. However, it is not completely consistent to
 include the terms of higher order in the calculations
 of the scattering amplitude, where only
 the lowest order lagrangian was used at the beginning. In
 any case, the ``right'' behavior of the total cross section, when the effects
of the baryon mass splitting are included, may indicate that by
 including the contributions from the terms of higher order, one can generate
a resonance at the right energy. This conclusion is also supported by the results
obtained in Ref. \cite {kzw} where the leading and next-to-leading terms
of the effective potential were used.
One notes, that the  calculations for the other possible channels 
are also in  reasonable agreement with the experimental data. They, however, cannot add
qualitatively new information about  the dynamics involved  
so we do not show the corresponding  results here.
One notes, to avoid confusion, that the phase space factors were,
of course, calculated with
the physical masses of the corresponding particles

Now a few remarks about the contributions involving $\eta$ mesons are in order.
The channels with $\eta$ mesons give moderate, but non-negligible 
contributions. In the case of low energy $K^{-}P$ interactions the tree level
amplitudes with $\eta$'s in the final state can only be off-shell by 
250-300 MeV, since there is not enough energy to generate the real $\eta\Lambda$
or $\eta\Sigma$ pairs. However, in  lowest order of HBChPT, when all baryons
 are assumed to be infinitely heavy, the degree of ``off-shellness'' becomes much less
(about 50 MeV) and negligible in lowest order. So we formally put the 
amplitudes on shell  for the processes like $K^{-}P \rightarrow \eta\Lambda$ and 
$K^{-}P \rightarrow \eta\Sigma$ and use the expression   
\begin{equation}
t_{K^{-}P \rightarrow \eta\Lambda(\Sigma)} \simeq {2\over f_{\pi}^2} E^{2}_i
\end{equation}
We note, however, that in the next orders, when the baryon mass difference  
should be taken into account, the contributions of the channels including $\eta$`s
may be considerably reduced, due to increased degree of the "off-shellness".
In fact, in Ref \cite{kzw} a good fit of the experimental data was obtained
without channels with $\eta$`s. So we conclude, that at the lowest order
its contribution is enhanced somewhat artificially. 

The other interesting physical quantities in low-energy KN scattering are 
the following branching ratios
\begin{eqnarray}
    \gamma &=& {{\Gamma(K^-p \rightarrow \pi^+ \Sigma^-)} \over
       {\Gamma(K^-p \rightarrow \pi^- \Sigma^+)}} = 2.36 \pm 0.04  \nonumber \\
    R_c &=& {{\Gamma(K^-p \rightarrow charged ~particles)} \over
       {\Gamma(K^-p \rightarrow all)}} = 0.664 \pm 0.011  \nonumber \\
    R_n &=& {{\Gamma(K^-p \rightarrow \pi^0 \Lambda)} \over
       {\Gamma(K^-p \rightarrow ~all~ neutral~ states)}} = 0.189 \pm 0.015
\end{eqnarray}

The experimental values are taken from Refs.\cite{nowak}
and \cite{tovee}. 
They should be compared with   the theoretical calculations,
which give the following numerical results 
\begin{equation}
\gamma=2.01, R_{c}=0.58, R_n=0.13.
\end{equation}
One can see that the  leading order terms give the dominant contribution,
 although higher order
corrections are apparently required.

One can see from Figs. 1-3 that, although the results of the calculations 
are in fairly good agreement with the experimental data, they are $extremely$
sensitive to the value of the cut-off parameter used. This means, that the loop
momenta in the vicinity of the cut-off may give a significant contribution.
Similar results were obtained and discussed in Ref.\cite {bpc} for the case
of $NN$ scattering. That, in turn, indicates that there is no small parameter,
providing convergence of the chiral expansion of the 
effective potential which is the sum of the tree level diagrams of  
increasing order, so that the chiral expansion cannot be arbitrarily truncated.
One could hope that taking into account the next-to-leading corrections to
the tree level amplitude might  decrease of the sensitivity of the calculations
to the value of the cut-off. However, in the calculations in the Ref.\cite{kzw}
it was demonstrated that inclusion of the next-to-leading terms does not significantly
affect the numerical value of the effective cut-off so the strong sensitivity 
to the value of  $\Lambda$ is still there  even if next-to-leading order corrections to the
tree level amplitude are included.
 One notes that, as long as the factorization property holds
at lowest order, one can consider the processes where the external momenta
are small compared to the cut-off. However, one can see that the factorization 
property breaks down when the terms of the next order are taken into account.
Then, in order to have a reliable scheme for the calculations one needs 
to estimate the effective potential in the region of momenta close to the value
of the cut-off, where, due to the absence of the small parameter of the 
order $m_\pi/\Lambda$, the chiral expansion cannot be applied.
In other words, using the cut-off introduces an additional length scale, which 
makes it rather difficult to implement the chiral expansion in the
standard sense. It is again worth mentioning that in the standard perturbative
procedure the dependence of the physical amplitudes on the cut-off can be
 removed order by order by using counterterms corresponding to the tree level
 terms of higher order. One notes, that in our case we have the power-law
divergencies, so the situation is again identical to that existing in  
nucleon-nucleon scattering, where the divergencies also have the power-law
character so that dimensional regularization and a cut-off scheme give different
results for the physical observables \cite{bpc}. So one can expect that
the same problem exists in the other systems, where the dominated dynamical
feature is the existence of the bound state near threshold. This, in turn,
means strong sensitivity to the short range effects similar to that found
for the case of NN scattering \cite{bpc}.

So we have to conclude that   the nonperturbative approach, described above,
should be considered as a rather phenomenological method to fit the
experimental data. The problem of the strong dependence of the physical amplitudes
on both the numerical parameters of the regulator and the way the regularization is done
will inevitably arise for the approaches of this type, when, due to the presence of
a subthreshold bound state, the chiral expansion is applied to the effective potential,
iterated then to infinite order. It seems, therefore  
that there is no way in this case to formulate an effective chiral theory
because of the absence of a small parameter in the chiral expansion of the
effective potential. One needs to stress that we have discussed one particular type
of a chiral approach. Effective field theory is based on the very general principles 
and should, therefore, work. The practical realization of those principles may, however,
be somewhat complicated due to existence of a subthreshold bound state.  
The similar conclusion was made in Ref. \cite{bpc}
for the case of NN scattering.

One notes that a possible way out was suggested in Ref. \cite{lee} where
the $\Lambda$(1405) resonance was treated as an explicit degree of freedom.
It was shown in \cite{bpc} that, once the $\Lambda$(1405) state is imbedded
in the lagrangian as an elementary field, the leading order calculations
can explain most of the branching ratios. A similar idea was demonstrated 
to be very successful  \cite{kap} for the  NN scattering. Since in this case
one can formulate a consistent chiral expansion with power counting rules,
this direction clearly deserves further study.

\section*{Acknowlegement}
Author would like to thank the Nuclear Theory Center at the Indiana University 
for the warm hospitality extended to him during his visit and Prof. E.Oset for
discussions.

\newpage
\begin{center}
{\Large FIGURES CAPTIONS}
\end{center}
\bigskip
FIG.1 The results of the theoretical calculations compared with experimental data \cite{nowak} 
for $K^{-}p$ elastic scattering. The upper curve corresponds to the calculations with  
$\Lambda$=580 MeV and the lower curve is the result obtained with $\Lambda$=600 MeV.
\vskip1cm
FIG.2  The same as in Fig.1 but for the $K^{-}p \rightarrow \Sigma^{0}\pi^{0}$ reaction. 
\vskip1cm
FIG.3 The results of the theoretical calculations for the $K^{-}p \rightarrow \Sigma^{-}\pi^{+}$
 reaction. The dashed line represents the results of the calculations when 
 the baryon mass difference is taken into account in the propagator $G_l$.
The solid curve is the result of the calculations in the strict ``heavy baryon'' limit. Both curves
are obtained with $\Lambda$=580 MeV.

\newpage
\section*{Appendix}

Here we present the results of the analytical calculation of the 
main value integral appearing when calculating the propagator
$G_l$. After the angular integration and using the standard expression
${1\over {a + i \epsilon}} =  {1\over {a}} - i\pi\delta(a)$ the mean value
integral can be defined as follows
\begin{equation}  
I= \int{dq q^2 {i\over {(k -\omega(q))\omega(q)}}},
\end{equation}  
where $\omega(q)$=$\sqrt {q^2+m^2}.$
The result of computation is $I= i(I_1 + I_2 + I_3)$,
 where
\begin{equation}
I_1= -k\ln(\Lambda + \sqrt {\Lambda^2+m^2}) +  
 (arctanh (\Lambda \sqrt {k^2 - m^2} 
 + m^2){1 \over {2k\sqrt {\Lambda^2+m^2}}})\sqrt {k^2 - m^2}, 
\end{equation}
\begin{equation}
I_2 = - (arctanh (-\Lambda \sqrt {k^2 - m^2} 
 + m^2){1 \over {2k\sqrt {\Lambda^2+m^2}}})\sqrt {k^2 - m^2} - \Lambda,
\end{equation}
\begin{equation}
I_3 =  k\ln(m) + \sqrt {-k^2 + m^2} arctanh {1 \over {-k^2 + m^2}}.
\end{equation}
 The calculation of the part with delta function
 is rather trivial so we do show the corresponding result here.

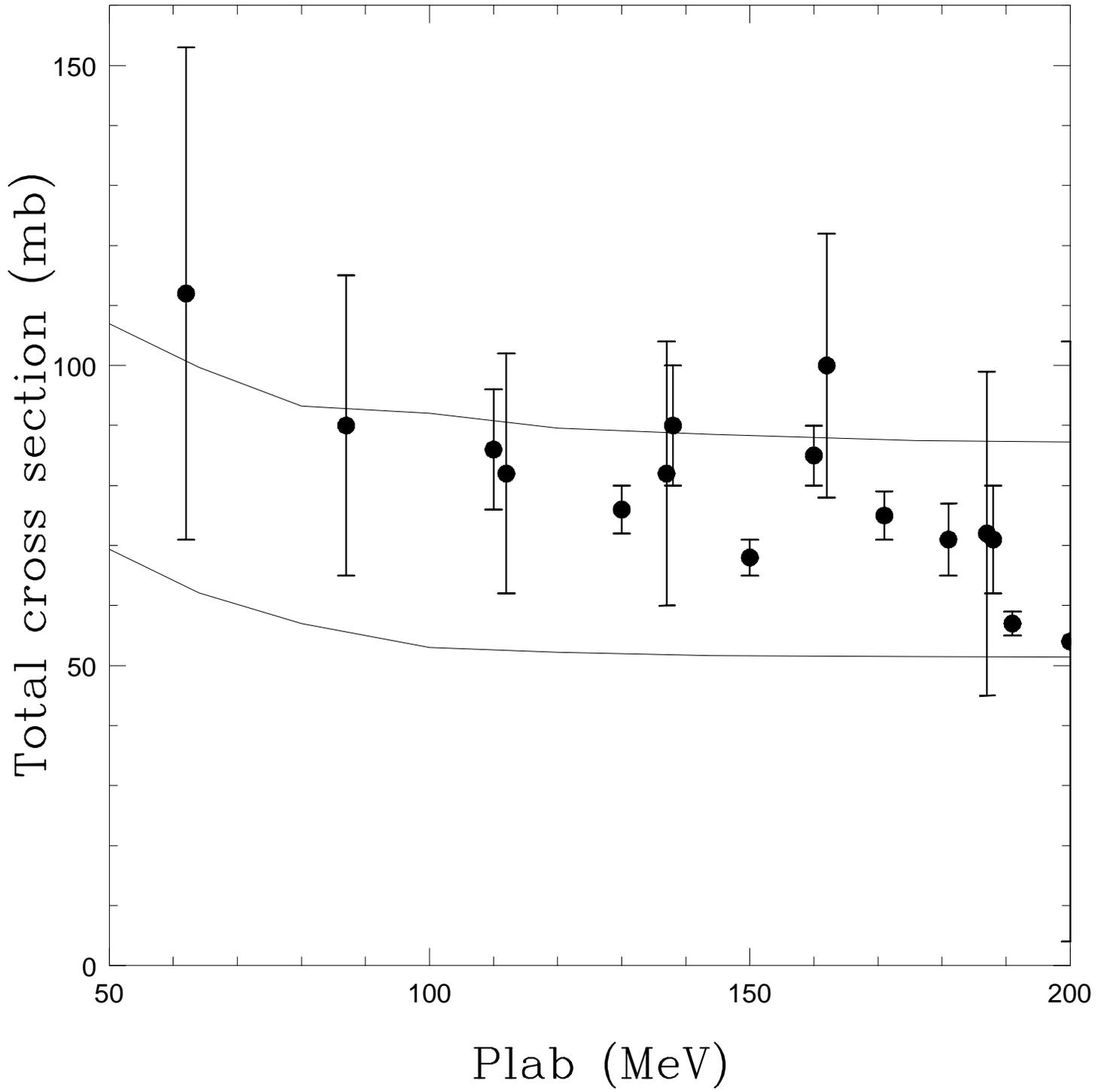
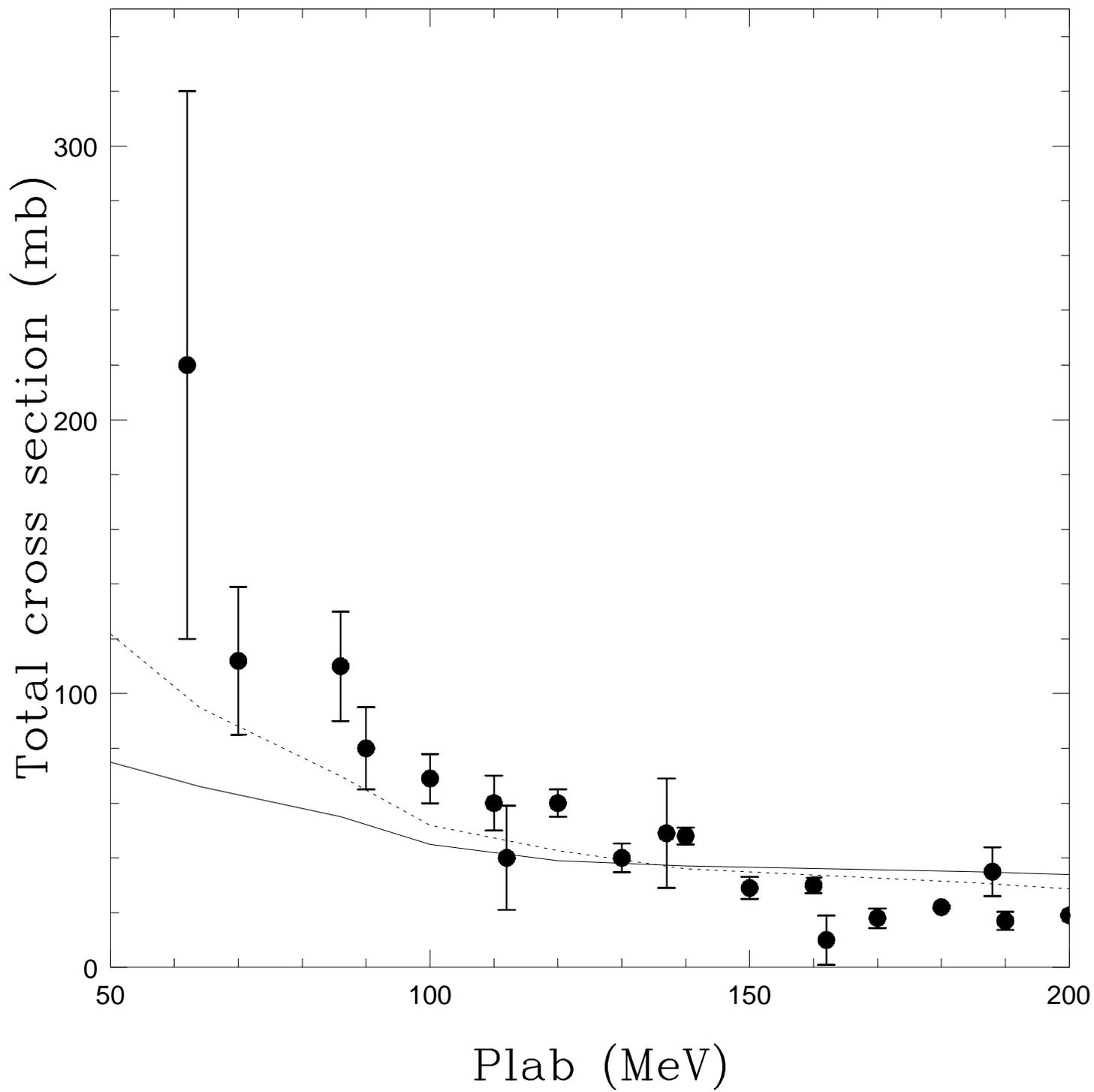
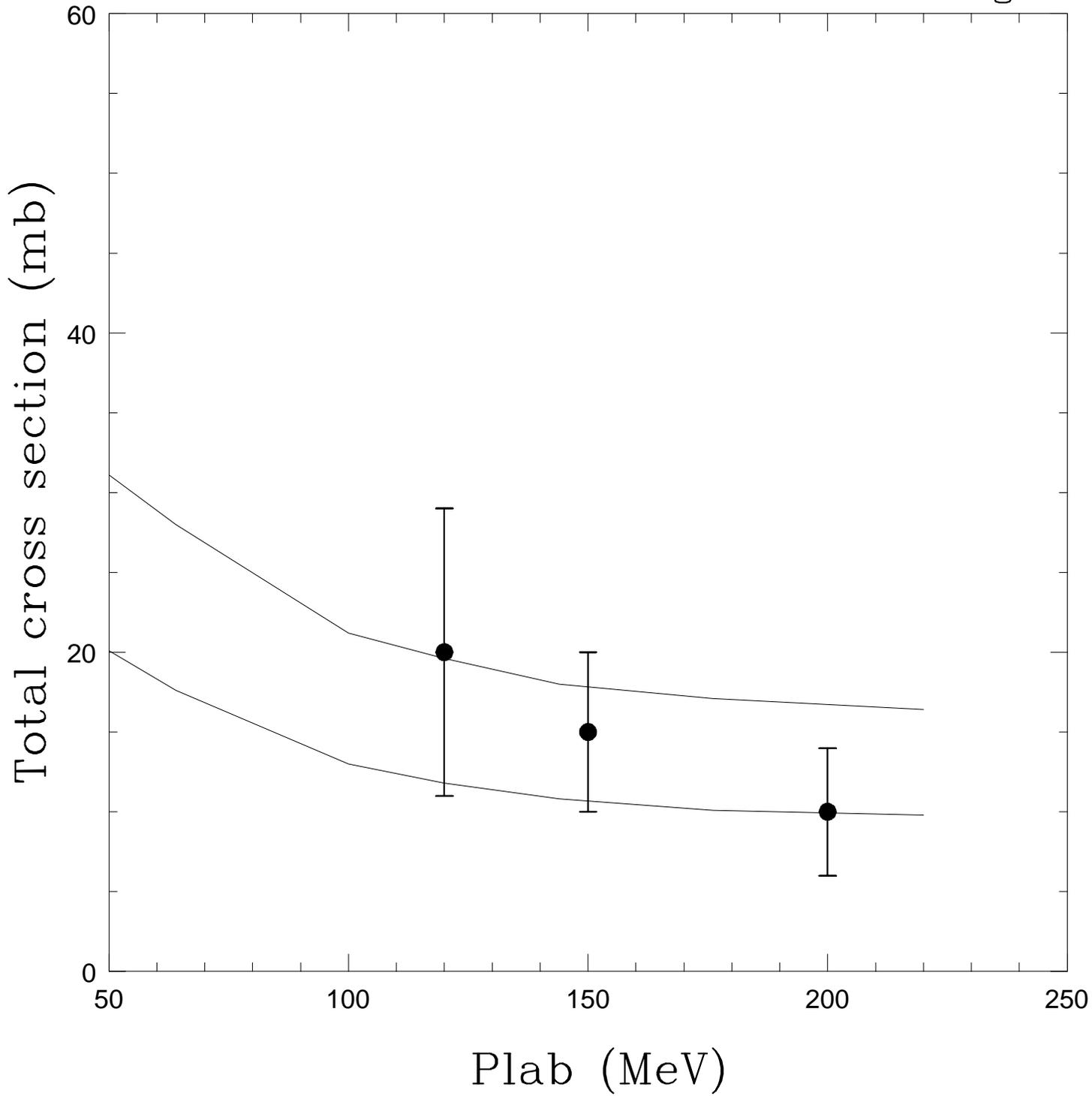
\end{document}